\documentstyle[aps,epsfig]{revtex}
\begin{document}
\draft
\title{Piezoelectric mechanism of orientation of a bilayer Wigner
crystal in a GaAs matrix}
\author{D.V.Fil}
\address{Institute for Single Crystals National Academy of Sciences of
Ukraine,
Lenin av. 60 Kharkov 61001 Ukraine\\
e-mail: fil@isc.kharkov.com}
\maketitle
\begin{abstract}
A mechanism for orientation of bilayer classical Wigner crystals in a
piezoelectric medium is considered. For the GaAs system the
piezoelectric correction to the electrostatic interaction between
electrons is calculated. It is shown that taking into account the
correction due to the piezoelectric effect leads to a dependence of the
total energy of the electron crystal on its orientation with respect to
the crystallographic axes of the surrounding matrix. A generalization
of Ewald's method is obtained for calculating the anisotropic
interaction between electrons in a Wigner crystal. The method is used
to calculate the energy of bilayer Wigner crystals in electron layers
parallel to the crystallographic planes (001), $(0{-}11)$, and (111) as
a function of their orientation and the distance between layers, and
the energetically most favorable orientation for all types of electron
lattices in a bilayer system is found. It is shown that phase
transitions between structures with different lattice symmetry in a
Wigner crystal can be accompanied by a change of its orientation.
\end{abstract}

\section{INTRODUCTION}

It is known that a system of electrons in the presence of a
neutralizing positive background at sufficiently low temperatures and
densities goes into a Wigner crystal state. In particular, this
situation takes place in two-dimensional electron layers in
AlGaAs--GaAs heterojunctions. The formation of a Wigner crystal phase
occurs under the condition that the average distance between electrons
is much greater than the effective Bohr radius (in the absence of an
external magnetic field) or the cyclotron radius (in high magnetic
fields). The latter situation corresponds to a filling factor $\nu  \ll
1$. In quantum Hall systems the formation of modulated electron
structures can occur in other regimes as well. For example, skyrmion
lattices can arise in the quantum Hall ferromagnet regime ($\nu \approx
1,1/3$).\cite{1} In low magnetic fields ($\nu \approx N+1/2$, where
$N$ is an integer and $N \geq 4)$ the formation of stripe structures
can occur at the top, partially filled Landau level. The
formation of such structures was predicted theoretically\cite{2} and
confirmed experimentally\cite{3} by observation of a strong
anisotropy of the conductivity.

An interesting question is that of the orientation of the nonuniform
electron structures relative to the crystallographic axes of the
surrounding matrix. In the two-dimensional electron layers realized in
AlGaAs heterostructures, an important influence on the orientation of
the electron lattice can be exerted by the piezoelectric interaction
between the electron and elastic subsystems. This possibility was first
pointed out in relation to Wigner crystals in Refs.\ \onlinecite{4} and
\onlinecite{5}. A piezoelectric mechanism for the orientation of the
stripe structure was considered in Ref.\ \onlinecite{6}. In particular,
in that paper an effect was detected wherein a reorientation of the
stripes in bilayer systems arises in the case when the period of the
structure is greater than the distance between layers. In Ref.\
\onlinecite{6} the electron subsystem was described using a
charge-density-wave model. The purpose of the present paper is to
examine the piezoelectric mechanism of orientation of nonuniform
electronic structures in bilayer systems in another limiting case,
corresponding not to a charge density wave but to a classical Wigner
crystal. As in Ref.\ \onlinecite{6}, we use a model which takes into
account the anisotropy of the elastic constants of the crystalline matrix.
We develop an approach whereby the energy of a system with an
anisotropic interaction between electrons can be calculated exactly;
this approach is a generalization of Ewald's method for calculating
Coulomb sums. As a particular case we obtain results pertaining to a
monolayer system. This topic was discussed previously in Ref.\
\onlinecite{5}, where consideration was limited to the model situation
of an isotropic crystal. Such a model does not answer the question of
the specific orientation that will be realized in the GaAs system, the
elastic properties of which are described by three elastic constants
rather than two. Furthermore, the method of rapidly convergent lattice
sums was not used in Ref.\ \onlinecite{5}. For the case of bilayer
systems, as far as we know, the question of a piezoelectric mechanism
of orientation of a Wigner crystal has not been considered before.

The lattice symmetry of a classical Wigner crystal is determined by the
minimum of its Coulomb energy. In a monolayer system the minimum is
achieved for a hexagonal lattice \cite{7}. In a bilayer system with the
same electron density in the two layers, five types of electron lattice
can form. The structure, dynamical properties, and melting criterion of
such systems have been studied in detail in Refs.\ \onlinecite{8,9,10}.
Quantum bilayer Wigner crystals in an external magnetic field were
treated in Refs.\ \onlinecite{11} and \onlinecite{12}. The possibility
of formation of bilayer Wigner crystals in superfluid helium films
was examined in Refs.\ \onlinecite{13} and \onlinecite{14}.

In a classical bilayer Wigner crystal the transition between
different crystalline phases is regulated by the parameters $\eta  =
d\sqrt{n}$, where $d$ is the distance between layers and $n$ is the
electron density in the layer. The case $\eta  = 0$ corresponds to a
monolayer system with a doubled density, and the case $\eta  = \infty $
to a system of two noninteracting layers. In both cases the minimum
energy corresponds to hexagonal lattices, with a period differing by a
factor of $\sqrt{2}$. Therefore, for finite $\eta $ there must be
transition phases: rectangular, square, and rhombic.

The piezoelectric interaction, generally speaking, can lead to a shift
of the boundaries between phases. In GaAs the piezoelectric interaction
is rather weak, and so the indicated effect will be small.
Nevertheless, since the Coulomb interaction in a system with cubic
lattice symmetry is isotropic, the piezoelectric correction to the
interaction between electrons can be important for determining the
orientation of the electron crystal.

\section{ENERGY OF THE PIEZOELECTRIC INTERACTION BETWEEN ELECTRONS
IN A WIGNER CRYSTAL}

Consider an infinite piezoelectric medium. The electrostatic potential
$\varphi $ of an electron placed at the origin of the coordinate system
is given by the following system of equations:
\begin{eqnarray}
{\rm div} {\bf D} = 4 \pi e \delta({\bf r}),\cr
{\partial  \sigma_{i k}\over \partial x_k}=0,
\label{1}
\end{eqnarray}
where
\begin{eqnarray}
D_i = -\varepsilon_{i k }{\partial \varphi \over \partial x_k}
- 4\pi \beta_{i, k l} u_{k l} \ \ -
\label{2}
\end{eqnarray}
is the electric displacement vector, and
\begin{eqnarray}
\sigma_{i k} =\lambda _{iklm} u_{lm} -\beta_{l,ik}
{\partial \varphi \over \partial x_l} \ \ -
\label{3}
\end{eqnarray}
is the stress tensor. Here $\varepsilon _{ik}$ is the dielectric
tensor, $\lambda _{iklm}$ is the tensor of elastic constants, $\beta
_{l,ik}$ is the tensor of piezoelectric moduli, and $u_{ik}$ is the
strain tensor. After a transformation to the Fourier components, the
system of equations (1) reduces to an algebraic system, and one can
easily write the solution for the electrostatic potential. Let us write
it out explicitly for a cubic system, the properties of which are
determined by three elastic constants $c_{11}$, $c_{12}$, and $c_{44}$,
one piezoelectric modulus $e_{14}$, and the dielectric constant $\varepsilon
$:
\begin{eqnarray}
\varphi_{\bf q} = {4 \pi e \over \varepsilon q^2 } -
{(4 \pi)^2 e \over \varepsilon q^2 } \chi {P(q_x,q_y,q_z)\over q^8
\rho^3 s_1^2({\bf q}) s_2^2({\bf q}) s_3^2({\bf q})} + O(\chi^2),
\label{4}
\end{eqnarray}
where $\chi  = e_{14}^{2}/\varepsilon c_{11}$ is a small parameter in
which the expansion is done, $s_{i}({\bf q})$ is the velocity of sound
with polarization $i$ in the direction ${\bf q}$, and $\rho $ is the
density of the medium. The function $P$ is a homogeneous 8th-degree
polynomial of the form
\begin{eqnarray}
P(q_x,q_y,q_z)= q^2 (a_1 q_x^2 q_y^2 q_z^2+ a_2 \sum_{l\ne k} q_l^4 q_k^2)+
a_3 \sum_{l\ne k} q_l^4 q_k^4,
\label{5}
\end{eqnarray}
where $l,k = x,y,z$, and
\begin{eqnarray}
a_1=c_{11}(2c_{12}^2- 2c_{11}c_{12}+c_{44}^2 - 2 c_{11}c_{44}),\cr
a_2=c_{11}^2 c_{44},\cr
a_3={1\over 2}c_{11}(c_{11}+c_{12})(c_{11}-c_{12}- 2 c_{44}).
\label{6}
\end{eqnarray}
The $x,y,z$ axes are directed along the fourfold axes of the
crystalline matrix. As we see from formula (4), the electrostatic
potential contains a correction $\delta \varphi _{{\bf q}}$, the
presence of which is due to the piezoelectric interaction. In an
isotropic crystal, in which the sound velocity is independent of the
direction and the coefficient $a_{3}$ in (5) is zero, the correction
linear in $\chi $ can be represented as an expansion in a finite number
of spherical harmonics:
\begin{eqnarray}
\delta\varphi_{\bf q} =-{(4 \pi)^2 e \over \varepsilon q^2 } \chi
\sum_n \sum_{m=-n}^n A_{n m} Y_{n m} (\Theta_{\bf q}, \psi_{\bf q}),
\label{7}
\end{eqnarray}
where $n$ is even and $n \leq 6$. In Eq.\ (7) $\psi _{{\bf q}}$ and
$\Theta _{{\bf q}}$ are the polar and azimuthal angles, respectively.
The coefficients $A_{nm}$ are expressed in terms of the longitudinal
and transverse sound velocities. Expansion (7) in the case of an
anisotropic crystal will also contain higher harmonics allowed by the
symmetry of the system. The coefficients $A_{nm}$ for the anisotropic
case can be found numerically. A calculation of these coefficients for
the GaAs system ($c_{11} = 12.3\times 10^{11}$ dyn/cm$^{2}$, $c_{12} =
5.7\times 10^{11}$ dyn/cm$^{2}$, $c_{44} = 6.0\times 10^{11}$
dyn/cm$^{2}$) shows that the main contribution to the sum in (7) is
given by the same harmonics ($n \leq 6)$ as in the isotropic system
(the coefficients of the higher harmonics are at least an order of
magnitude smaller). It should be noted that, because the relations
between the expansion coefficients are determined by three rather than
two elastic constants, the problem cannot be reduced to isotropic even by
limiting consideration to the lowest harmonics.

\begin{center}
\begin{figure}
\centerline{\epsfig{figure=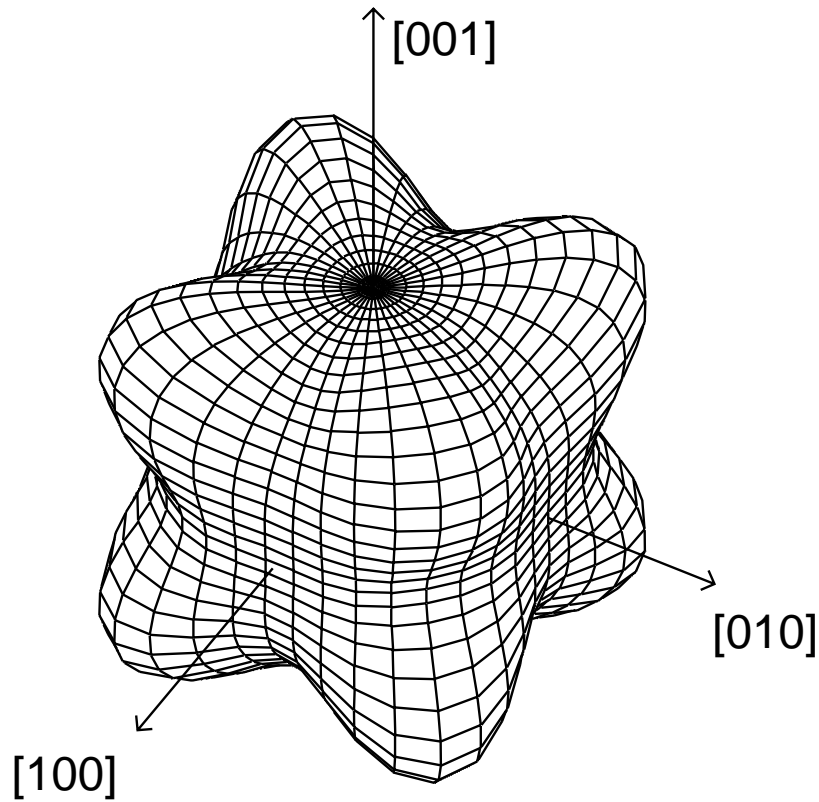,width=12cm}}
\vspace{0.5cm}
\caption{
The anisotropy of piezoelectric interaction between electrons in
a GaAs host matrix.}
\label{f1}
\end{figure}
\end{center}

Using expansion (7), one can easily find the correction to the
electrostatic interaction between electrons in a piezoelectric medium.
After doing the inverse Fourier transformation we find
\begin{eqnarray}
\delta V({\bf r})= - { e^2\chi\over \varepsilon r}
G(\Theta_{\bf r},\psi_{\bf r}),
\label{8}
\end{eqnarray}
where
\begin{eqnarray}
G(\Theta_{\bf r},\psi_{\bf r}) =
4 \pi \sum_{nm} A_{nm}(-1)^{n\over 2}
{n!\over 2^n [(n/2)!]^2} Y_{nm}(\Theta_{\bf r},\psi_{\bf r}).
\label{9}
\end{eqnarray}
The form of the function $G$ calculated for the GaAs crystal is shown
in Fig.\ \ref{f1} (the coordinate axes are chosen along the fourfold
axes). As we see from Eq.\ (8), the interaction between electrons
contains a contribution corresponding to attraction, the strength of
which depends on the direction of ${\bf r}$ and decays as $1/r$. Since
this decay is as slow as that of the Coulomb interaction, the
convergence of the lattice sums will be slow. Therefore, for correct
calculation of the piezoelectric correction to the energy of a Wigner
crystal it is better use a method of rapidly convergent sums, modified to
take into account the anisotropy of the interaction.

Consider a bilayer electron system placed in an infinite
piezoelectric medium and oriented normal to a certain crystalline direction.
In this case it is convenient to make a change of variable in Eq.\ (9)
to the new angles $\Theta $ and $\psi $, measured from the normal to
the surface of the electron layer and from an axis lying in the plane
of the layer, respectively (the overall structure of expression (9) is
preserved --- only the values of the expansion coefficients are changed).
If the structure of the electron crystal is assumed fixed, then its
orientation is determined by the contribution of the terms in (9) which
depend on the angle $\psi _{{\bf r}}$ in the chosen reference frame.
The latter correspond to harmonics with $m\neq 0$. For calculating the
lattice sums we rewrite the dependence of the $\psi _{{\bf
r}}$-dependent part of the interaction between electrons in the form
\begin{eqnarray}
V_{an}(r,\Theta_{\bf r},\psi_{\bf r})= - {e^2 \chi\over \varepsilon r}
\sum_{l\ge 0}\sum_{|m|>0}
B_{lm} \cos^{l}\Theta_{\bf r} \sin^{|m|}\Theta_{\bf r}
e^{i m \psi_{\bf r}},
\label{10}
\end{eqnarray}
Using the explicit form of the spherical harmonics, we can express
the coefficients $B_{lm}$ in terms of the coefficients $A_{nm}$. Since
the index $n$ in Eqs.\ (7) and (9) takes on only even values, the
coefficients $B_{lm}$ are nonzero only for $l$ and $m$ having the same
parity. If we keep a finite number of spherical harmonics in expansion
(9), then the sum in (10) will also contain a finite number of terms.
We note that to preserve the point symmetry with respect to the angle
$\psi $ when a finite number of harmonics is taken into account, the
transformation to the indicated coordinate system should be done in
Eq.\ (4) and then the values of the expansion coefficients in (7)
should be found numerically.

Taking Eq.\ (10) into account, we write the anisotropic contribution to
the energy of the electron crystal in the form
\begin{eqnarray}
E_{an}=E_{an}^{in}+E_{an}^{out}+E^{BG}_{an},
\label{11}
\end{eqnarray}
where
\begin{eqnarray}
E_{an}^{in}= -{e^2 \chi\over \varepsilon } \sum_{|m|>0} B_{0m}
\sum_{{\bf R}\ne {\bf R}^{'}}
{1\over |{\bf R}-{\bf R}{'}|}
e^{i m \psi_{{\bf R}^{'}-{\bf R}}}
\label{12}
\end{eqnarray}
corresponds to the contribution of the interaction within the
layers,
\begin{eqnarray}
E_{an}^{out}=
- {e^2 \chi\over \varepsilon} \sum_{l\ge 0}\sum_{|m|>0}
 B_{lm}
\sum_{{\bf R},{\bf R}^{'}}
{ d^{l} |{\bf R}-{\bf R}^{'}-{\bf c}|^{|m|} \over
 [|{\bf R}-{\bf R}^{'}-{\bf c}|^2+d^2]^{(|m|+l+1)/2}}
e^{i m \psi_{{\bf R}^{'}+{\bf c}-{\bf R}}}
\label{13}
\end{eqnarray}
describes the contribution of the interlayer interaction, and $E_{\rm
an}^{BG}$ gives the correction to the interaction with the positive
neutralizing background. The vectors ${\bf R}$ and ${\bf R'}$ in Eqs.\
(12) and (13) are lattice vectors, and the vector ${\bf c}$ specifies
the displacement of the upper sublattice relative to the lower. The
lattice sums in (12) and (13) can be reduced to a rapidly convergent
form with the use of a modified version of Ewald's method (see
Appendix). Employing this method gives
\begin{eqnarray}
E_{an}=-{N e^2 \chi\over \varepsilon } \sqrt{n} (S_{in}+S_{out}),
\label{14}
\end{eqnarray}
where $N$ is the total number of particles in the layer, and
\begin{eqnarray}
S_{in} =
\sum_m B_{0 m}
\{\sum_{{\bf R}\ne 0} e^{i m \psi_{\bf R}} \Phi(m,\pi n R^2) +
i^{|m|} \sum_{{\bf G}\ne 0}
e^{i m \psi_{\bf G}} \Phi(m,{G^2\over 4 \pi n})\},
\label{15}
\end{eqnarray}
\begin{eqnarray}
S_{out} =
\sum_{l m}
B_{lm}
 \{ \sum_{\bf R}
{ d^l |{\bf R}+{\bf c}|^{|m|} \over
 [|{\bf R}+{\bf c}|^2+d^2]^{|m|+l\over 2}}
e^{i m \psi_{{\bf R}+{\bf c}}}
\Phi\left( l+|m|,\pi n [|{\bf R}+{\bf c}|^2+d^2]\right)  \cr+
i^{|m|} \sum_{{\bf G}\ne 0} e^{-i{\bf G c}+i m \psi_{\bf G}}
\Psi(l,m, {G^2\over 4 \pi n}, \pi n d^2) \}
\label{16}
\end{eqnarray}
where ${\bf G}$ are reciprocal lattice vectors. In Eqs.\ (15) and
(16) we have introduced the functions
\begin{equation}
\Phi(m,x)=\sqrt{\pi \over x} {\Gamma({|m|+1\over 2},x)\over
\Gamma({|m|+1\over 2})}
\label{17}
\end{equation}
\begin{eqnarray}
\Psi(l,m,x,y)={1\over 2} \sqrt{\pi \over x} {1\over \Gamma({l+|m|+1\over2}) }
\sum_{s=0}^{N(l,m)} C_{N(l,m)+s}^{2s} (xy)^{|m|+l-2 s\over 4}
\cr\times [e^{-2\sqrt{xy}} F(s,\sqrt{x}-\sqrt{y})+
(-1)^{l+|m|-2 s\over 2} e^{2\sqrt{xy}} F(s,\sqrt{x}+\sqrt{y})]
\label{18}
\end{eqnarray}
In Eq.\ (18) the $C_{i}^{j}$ are binomial coefficients,
$N(l,m) = \max[(|m|-l)/2,(l-|m|-2)/2]$, and
\begin{eqnarray}
F(s,z)=\Gamma(s+1/2)-{\rm sgn}(z)\gamma(s+1/2, z^2).
\label{19}
\end{eqnarray}
In Eqs.\ (17)--(19) $\Gamma (x)$ is the gamma function, and $\Gamma
(k,x)$ and $\gamma (k,x)$ are incomplete gamma functions. We note that
for $l,m$ equal to zero (these terms have not been taken into account,
since they give a direction-independent correction to the
interaction) the sums (15) and (16) reduce to the known expressions for
the isotropic case.\cite{7,8}

\section{ORIENTATION OF A BILAYER WIGNER CRYSTAL IN A
G\lowercase{a}A\lowercase{s} MATRIX}

Let us use the results of the previous Section to determine the
orientation of bilayer Wigner crystals lying in the planes (001),
$(0{-}11)$, and (111) in the GaAs matrix. In describing the
piezoelectric interaction with allowance for anisotropy of the elastic
constants, we keep in expansion (7) only the harmonics with $n \leq 18$ and
$|m| \leq 12$. We note that in the cases considered below, harmonics
with $n>6$ only influence the orientation of hexagonal structures in
layers parallel to the (001) plane and that of square structures in layers
parallel to the (111) plane. In these cases the symmetry of the system
leads to the vanishing of the contribution of the lower harmonics to
the energy of the Wigner crystal.

The structure of a Wigner crystal in a bilayer system is specified by
the primitive lattice vectors ${\bf R}_{1}$ and ${\bf R}_{2}$ and the
vector ${\bf c}$ of the relative displacement of the sublattices in
adjacent layers. The values of these vectors for the five types of
lattice considered are listed in Table \ref{t1}. The values of $\eta $
for which the change of lattice symmetry occurs were obtained in Ref.\
\onlinecite{8}. Since we will have need of the functions $\delta (\eta
)$ (for a rectangular lattice) and $\alpha (\eta )$ (for a rhombic
lattice), we have repeated the calculations of Ref.\ \onlinecite{8}.
According to the results of our calculations, the transition between
the rectangular and square phases occurs at $\eta \approx 0.263$, that
between the square and rhombic phases at $\eta \approx 0.621$, and that
between the rhombic and double hexagonal at $\eta \approx 0.732$. The
first two are second-order transitions, and the last is first-order.
These results reproduce those of Ref.\ \onlinecite{8}. (We will not
analyze the transition between the simple hexagonal and rectangular
phases, which, according to Ref.\ \onlinecite{8}, occurs at $\eta  =
0.006$, since at this transition the orientation of the electron
crystal changes only slightly.)

\begin{table}
\caption{Structure of double layer Wigner crystals.
${\bf R}_1$, ${\bf R}_2$ are the primitive vectors of the direct lattice;
${\bf G}_1$, ${\bf G}_2$, the primitive vectors of the reciprocal lattice;
${\bf c}$, the vector of relative displacement of sublattices in adjacent
layers; $n$, the one layer electron density.
}
\label{t1}
\begin{tabular}{lccccccc}
Lattice type &
${\bf R}_1$&
${\bf R}_2$&
${\bf G}_1$&
${\bf G}_2$&
${\bf c}$&
Variable parameter&n\\
\tableline \\
simple hexagonal&$(a,0)$&$(0,\sqrt{3} a)$&
$({2\pi \over a},0)$&$(0,{2\pi \over a\sqrt{3}})$&
${a\over 2}(1,\sqrt{3})$&-&${1\over a^2\sqrt{3}}$\\
 \\
rectangular&$(a,0)$&$(0,a\delta )$&
$({2\pi \over a},0)$&$(0,{2\pi \over a\delta })$&
${a\over 2}(1,\delta )$&$1<\delta<\sqrt{3}$&${1\over a^2\delta }$\\
 \\
square&$(a,0)$&$(0,a )$&
$({2\pi \over a},0)$&$(0,{2\pi \over a })$&
${a\over 2}(1,1 )$&-&${1\over a^2}$\\
 \\
rhombic&$(a,0)$&$a(\cos\alpha ,\sin\alpha  )$&
${2\pi \over a}(1,-\cot \alpha )$&
$(0,{2\pi \over a\sin \alpha  })$&
${a\over 2}(1+\cos\alpha ,\sin \alpha )$&${\pi \over 3}<\alpha
<{\pi \over 2}$&${1\over a^2 \sin \alpha  }$\\
 \\
double hexagonal&$(a,0)$&$({a\over 2},{a\sqrt{3}\over 2})$&
${2\pi \over a}(1,-{1\over \sqrt{3}})$&
$(0,{4\pi \over a\sqrt{3}  })$&
${a\over 2}(1 ,{1\over \sqrt{3}} )$&-&
${2\over a^2 \sqrt{3}  }$
\end{tabular}
\end{table}

The anisotropic contribution to the energy of a bilayer electron
crystal as a function of its orientation in the (001), $(0{-}11)$, and
(111) planes is shown in Figs.\ \ref{f2}, \ref{f3}, and {\ref{f4} for
various values of the parameter $\eta $ (the energy is given as the
energy per electron in units of $e^{2}\chi \sqrt{n}/2\varepsilon )$.

For a two-dimensional Wigner crystal lying in the (001) plane, the
anisotropic contribution to the energy of the simple hexagonal
structure does not exceed $2\times 10^{-2}$ (in the indicated units). A
minimum of the energy is reached at an angle between ${\bf R}_{1}$ and
the [100] axis which is a multiple of 30$^{\circ}$. The absolute
minimum of the energy of the rectangular phase occurs when one of the
primitive lattice vectors is directed along the [100] axis. At small
$\eta $ there are also local minima corresponding to angles $\beta
\approx \pm 30^{\circ}$ between ${\bf R}_{1}$ and one of the fourfold
axes. With increasing $\eta $ the local minima vanish, and the
anisotropy energy increases substantially. A minimum of the energy of
the square phase corresponds to an orientation of the elementary
lattice vectors along the fourfold axes. The vector ${\bf c}$ here is
oriented along one of the twofold axes. This orientation of the vector
${\bf c}$ is preserved in the rhombic phase as well (the vectors ${\bf
R}_{1}$ and ${\bf R}_{2}$ rotate smoothly on changes in $\eta $). The
transition to the hexagonal phase as the parameter $\eta $ is increased
further is accompanied by a jumplike change in the orientation of the
vectors ${\bf R}_{1}$ and ${\bf c}$ and a sharp decrease in the
anisotropy energy.
\begin{center}
\begin{figure}
\centerline{\epsfig{figure=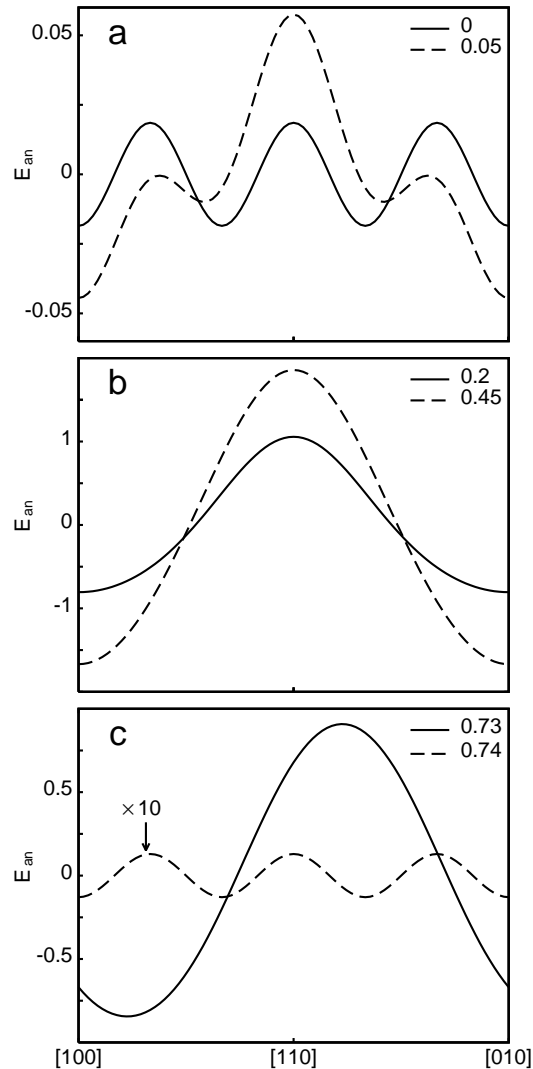,width=8cm}}
\vspace{0.5cm}
\caption{
The dependence of the anisotropic part of the piezoelectric
correction to the energy of a double layer Wigner crystal in
a (001) plane versus the direction  of
the ${\bf R}_1$ vector for the values of  $\eta$ specified;
a, a simple hexagonal phase and a rectangular phase at small $\eta$;
b, a rectrangular phase at large
$\eta$ and a square phase;
c, a rhombic and double hexagonal phases near the first-order
transition point.
The energy is given in units of
$e^2 \chi \sqrt{n}/2\varepsilon $.}
\label{f2}
\end{figure}
\end{center}

In the case of two-dimensional electron layers parallel to the
$(0{-}11)$ plane, the calculation gives the following results. A
minimum of the energy of the simple hexagonal structure is reached when one
of the primitive lattice vectors is oriented along the [100] axis.
When the distance between layers is increased and the rectangular
structure is formed, there arise two distinct equilibrium orientations, one of
which corresponds to a local minimum (${\bf R}_{1}$ directed along the
[100] axis) and the other to a global minimum (${\bf
R}_{1}$ directed at an angle $\beta \approx 60^{\circ}$ to the [100] or
$[{-}100]$ axis). Near the points of transition to the square phase the
local minimum vanishes, and a rapid reorientation of the electron
lattice occurs. The energy of the square phase is minimum when one of
the primitive lattice vectors is directed at an angle $\beta  =
45^{\circ}$ to the [100] axis, i.e., to the vector ${\bf c}$, which is
parallel to the [100] or the [011] axis. After the transition to the
rhombic phase the direction of the vector ${\bf c}$ parallel to the
[011] axis corresponds to the global minimum, while the direction
parallel to the [100] axis corresponds to a local minimum. As $\eta $ is
increased, the local minimum splits into two, corresponding to a
direction of the vector ${\bf c}$ at an angle $\pm \beta $ ($\beta
<30^{\circ})$ to this axis. At the transition to the double hexagonal
phase all three minima become equal (the orientation of ${\bf R}_{1}$
changes abruptly at this transition).
\begin{center}
\begin{figure}
\centerline{\epsfig{figure=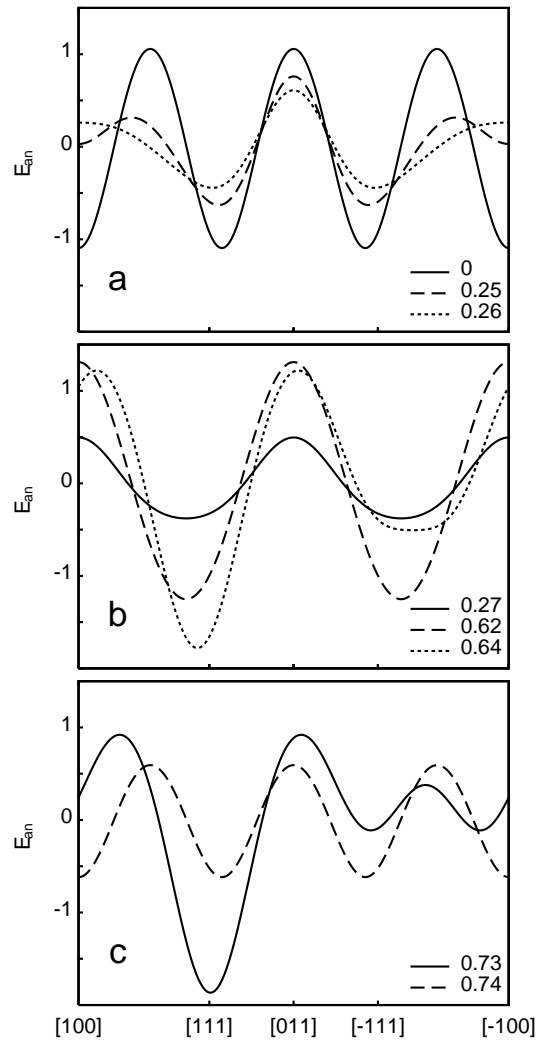,width=8cm}}
\vspace{0.5cm}
\caption{
The anisotropic part of the
energy of the  Wigner crystal in
a (0-11) plane versus the direction  of
the ${\bf R}_1$ vector for the values of  $\eta$ specified;
a, a simple hexagonal  and a rectangular phases;
b, a square and rhombic phases near the second-order transition point;
c, a rhombic and double hexagonal phases near the first-order
transition point.}
\label{f3}
\end{figure}
\end{center}

In a bilayer structure lying parallel to the (111) plane the minimum
of the energy of the simple hexagonal and rectangular phases correspond
to a direction of ${\bf R}_{1}$ at an angle $\beta  = 30^{\circ}$ to
one of the twofold axes lying in the (111) plane. As the point of
transition to the square phase is approached, the anisotropic
contribution decreases sharply. A sharp reorientation occurs near the
transition point. For the square phase a minimum of the energy is
reached when the vector ${\bf R}_{1}$ is directed at an angle $\beta  =
\pm 15^{\circ}$ to one of the twofold axes. At the transition to the
rhombic phase the anisotropy again increases. The energy of the rhombic
phase is minimum in the case when the vector ${\bf c}$ is oriented
along one of the twofold axes. At the transition to the double
hexagonal phase the orientation of the vector ${\bf c}$ changes
abruptly --- it deviates somewhat from the twofold axis. We note that
in the double hexagonal phase the energetically most favorable
orientation of the lattice vectors differs from the case of the simple
hexagonal phase. The latter is due to the absence of a center of
inversion in the double hexagonal structure.

\begin{center}
\begin{figure}
\centerline{\epsfig{figure=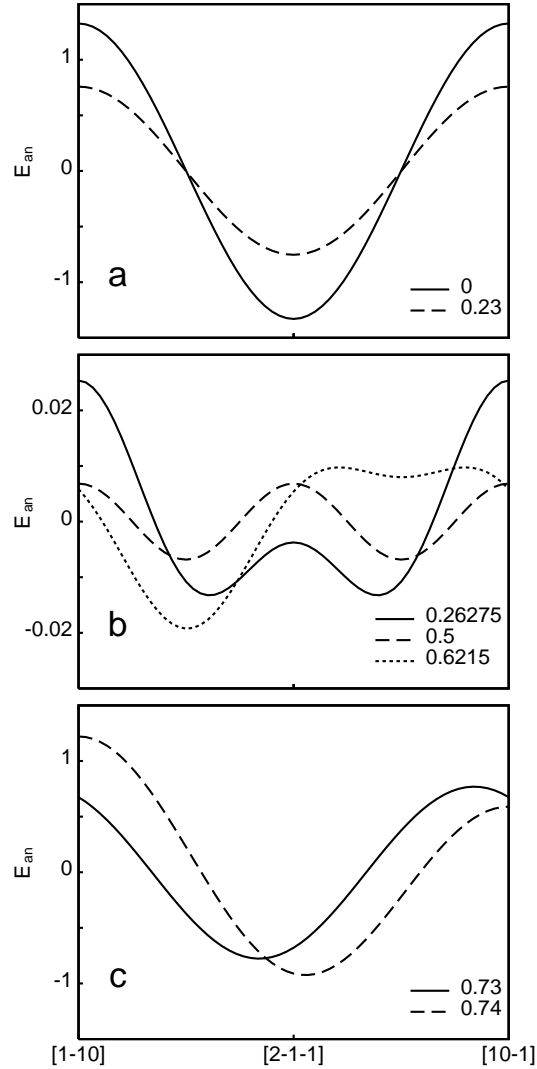,width=8cm}}
\vspace{0.5cm}
\caption{
The anisotropic part of the
energy of the  Wigner crystal in
a (111) plane versus the direction  of
the ${\bf R}_1$ vector for the values of  $\eta$ specified;
a, a simple hexagonal  and a rectangular phases;
b, a rectangular and
and a rhombic phases near the second-order transition point and a square
phase;
c, a rhombic and double hexagonal phases near
the first-order transition point.}
\label{f4}
\end{figure}
\end{center}

The results show that the orientation of a bilayer Wigner crystal in
a piezoelectric matrix is determined by the plane in which the
electron layer is located, the type of electron lattice, and the value
of the parameter $\eta $. Interestingly, the abrupt (jumplike)
character of the reorientation is observed not only at the transition
from the rhombic to the double hexagonal phase (this is the
expected effect, since it accompanies a first-order transition). A very
rapid reorientation also takes place at the second-order transition
from the rectangular to the square phase.

The value of the anisotropy energy is determined by the parameter $\chi
$, which in GaAs is of the order of $2\times 10^{-4}$. The typical
difference between the Coulomb energy in the different phases is of the
order of $10^{-2}e^{2}\sqrt{n}/\varepsilon $ per electron,\cite{7,8}
i.e., according to the results reported, the piezoelectric interaction
in the system under study is approximately two orders of magnitude
smaller; hence, it has a weak influence on the phase diagram and
only determines the orientation of the electron crystal. Nevertheless,
in other systems in which the value of the piezoelectric modulus is
larger, one can expect a radical rearrangement of the phase diagram.
Analogous effects for a monolayer system were discussed in Ref.\
\onlinecite{4}. The approach considered in the present paper makes it
possible to investigate this possibility in detail for the case of
bilayer electron crystals.

\section*{APPENDIX}

Let us transform to the rapidly convergent form of the expression
\begin{eqnarray}
S_{\pm m}=\sum_{{\bf R}\ne 0} { e^{\pm i m \psi_{\bf R}} \over R},
\eqnum{A.1}
\end{eqnarray}
where $m>0$. We introduce the function
\begin{eqnarray}
T_{\pm m}({\bf r},{\bf q})=
e^{-i {\bf q r}}\sum_{|{\bf r}+{\bf R}|}
{e^{i {\bf q} ({\bf R}+{\bf r}) \pm i m \psi_{{\bf r}+{\bf R}}}
\over |{\bf r}+{\bf R}|}
 - {e^{\pm i m \psi_{\bf r}}\over r}
\eqnum{A.2}
\end{eqnarray}
such that
\begin{eqnarray}
S_{\pm m}= \lim_{{\bf r}\to 0,{\bf q}\to 0} T_{\pm m}({\bf r},{\bf q}).
\eqnum{A.3}
\end{eqnarray}
We use the identity
\begin{eqnarray}
{{\gamma ({m+1\over 2},x)+
\Gamma({m+1\over 2},x)}\over \Gamma ({m+1\over 2})}\equiv 1,
\eqnum{A.4}
\end{eqnarray}
With allowance for (A.4) the quantity $T_{\pm m}$ can be written in the
form a sum
\begin{eqnarray}
T_{\pm m}({\bf r},{\bf q})=T_{\pm m,1}({\bf r},{\bf q})
+T_{\pm m,2}({\bf r},{\bf q}),
\eqnum{A.5}
\end{eqnarray}
where
\begin{eqnarray}
T_{\pm m,1}({\bf r},{\bf q})=
\sum_{{\bf R}\ne 0} {e^{i {\bf q R}\pm i m \psi_{{\bf r}+{\bf R}}}
\over |{\bf r}+{\bf R}|}
{\Gamma({m+1\over 2},\pi n  |{\bf r}+{\bf R}|^2)
\over \Gamma ({m+1\over 2})} -
{ e^{\pm i m \psi_{\bf r}}\over r}
{\gamma ({m+1\over 2},\pi n r^2 )\over \Gamma ({m+1\over 2})},
\eqnum{A.6}
\end{eqnarray}
\begin{eqnarray}
T_{\pm m, 2}({\bf r},{\bf q})=
e^{-i {\bf q r}}\sum_{{\bf R}}
{e^{i {\bf q} ({\bf r}+{\bf R})\pm i m
\psi_{{\bf r}+{\bf R}}}\over |{\bf r}+{\bf R}|}
{\gamma({m+1\over 2},\pi n |{\bf r}+{\bf R}|^2)
\over \Gamma ({m+1\over 2})}.
\eqnum{A.7}
\end{eqnarray}
We note that the last term in (A.6) vanishes in the limit ${\bf
r}\to 0$.

For the transformation $T_{\pm m,2}$ we substitute the definition of
the function $\gamma (\alpha ,x)$ into (A.7):
\begin{eqnarray}
T_{\pm m,2}({\bf r},{\bf q})={2\over \Gamma({m+1\over 2}) }
e^{-i {\bf q r}}
\int_0^{\sqrt{\pi n}} d \xi \xi^{m} \sum_{\bf R}
|{\bf r}+{\bf R}|^{m} e^{i {\bf q} ({\bf r}+{\bf R})
 \pm i m \psi_{{\bf r}+{\bf R}}
-\xi^2 |{\bf r}+{\bf R}|^2}.
\eqnum{A.8}
\end{eqnarray}
Expanding (A.8) in a Fourier series in the reciprocal lattice vectors,
we get
\begin{eqnarray}
T_{\pm m,2}({\bf r},{\bf q})
={2 n\over \Gamma({m+1\over 2}) } \sum_{\bf G}
 e^{-i ({\bf q}+{\bf G}) {\bf r}}
\int_0^{\sqrt{\pi n}} d \xi \xi^{m} \int d^2 \mbox{\boldmath$\rho $}
\rho ^{m} \exp [i \mbox{\boldmath$\rho $}
({\bf q}+{\bf G})\pm i m \psi_{\mbox{\boldmath$\rho $}}
 -\xi^2 \rho ^2].
\eqnum{A.9}
\end{eqnarray}
Evaluating the integral over $\mbox{\boldmath$\rho $}$, we find
\begin{eqnarray}
T_{\pm m,2}({\bf r},{\bf q})
=i^m {2\pi n\over \Gamma({m+1\over 2}) } \sum_{\bf G}
e^{-i ({\bf q}+{\bf G}) {\bf r}
\pm i m \psi_{{\bf q}+{\bf G}}} \left({|{\bf q}+{\bf G}|\over 2}
\right)^{m}
\int_0^{\sqrt{\pi n}} d \xi {1\over \xi^{m+2}}
\exp(-{|{\bf q}+{\bf G}|^2\over 4 \xi^2}).
\eqnum{A.10}
\end{eqnarray}
Making the change of variables $\xi  = |{\bf q}+{\bf G}|/2t$, we arrive
at the form
\begin{eqnarray}
T_{\pm m,2}({\bf r},{\bf q})
=i^m{2\pi n\over \Gamma({m+1\over 2}) }
\sum_{\bf G} e^{-i ({\bf q}+{\bf G}) {\bf r}
\pm i m \psi_{{\bf q}+{\bf G}}} {1\over {|{\bf q}+{\bf G}|}}
\Gamma({m+1\over 2}, {|{\bf q}+{\bf G}|^2\over {4\pi n}}).
\eqnum{A.11}
\end{eqnarray}
Substituting formulas (A.1), (A.3), (A.7), and (A.11) into Eq.\ (12) and
introducing the function (17), we arrive at Eq.\ (15). The term with
${\bf G} = 0$ in (A.11) drops out, since it cancels with the intralayer
interaction with the positive neutralizing background that appears in
$E_{\rm an}^{BG}$.

An analogous transformation may be done for the sum
\begin{eqnarray}
S_{l,\pm m}=\sum_{{\bf R}} {d^{l} |{\bf R}+{\bf c}|^{m}
e^{\pm i m \psi_{{\bf R}+{\bf c}}}
\over (|{\bf R}+{\bf c}|^2+d^2)^{(l+m+1)/2}}.
\eqnum{A.12}
\end{eqnarray}

Here we also use the identity (A.4), with $m$ replaced by $m+l$. We get
\begin{eqnarray}
S_{l,\pm m} = T_{l,\pm m,1}(0,0)+\lim_{{\bf r}\to 0,{\bf q}\to 0}
T_{l,\pm m,2}({\bf r},{\bf q}),
\eqnum{A.13}
\end{eqnarray}
where
\begin{eqnarray}
T_{l,\pm m, 1}(0,0)=
\sum_{\bf R}
{d^{l} |{\bf R}+{\bf c}|^{ m}e^{\pm i m \psi_{{\bf R}+{\bf c}}}
\over (|{\bf R}+{\bf c}|^2+d^2)^{(l+m+1)/2}}
 {\Gamma[{l+m+1\over 2}, \pi n (|{\bf R}+{\bf c}|^2+d^2)]
\over \Gamma({l+m+1\over 2}) }.
\eqnum{A.14}
\end{eqnarray}

The quantity $T_{l,\pm m,2}({\bf r},{\bf q})$, after we change to
summation over ${\bf G}$ and do the integration over
$\mbox{\boldmath$\rho $}$, reduces to the form
\begin{eqnarray}
T_{l,\pm m,2}({\bf r},{\bf q})=i^m {2 \pi n\over
\Gamma({l+m+1\over 2}) }\sum_{{\bf G}}
e^{- i {\bf G c}-i({\bf q}+{\bf G}) {\bf r}\pm
i m \psi_{{\bf q}+{\bf G}}} d^{l}
\left( {|{\bf q}+{\bf G}|\over 2}\right) ^{m} \cr \times
\int_0^{\sqrt{\pi n}} d\xi \xi^{l-m-2}
\exp\left( -\xi^2 d^2 - {|{\bf q}+{\bf G}|^2\over 4 \xi^2}\right).
\eqnum{A.15}
\end{eqnarray}
Evaluation of the integral in (A.15) gives
\begin{eqnarray}
T_{l,\pm m,2}({\bf r},{\bf q})
=i^m {\pi n\over \Gamma({l+m+1\over 2}) }\sum_{{\bf G}}
e^{- i {\bf G c}-i({\bf q}+{\bf G}) {\bf r}
\pm i m \psi_{{\bf q}+{\bf G}}}
{1\over |{\bf q}+{\bf G}|}
\sum_{s=0}^{N(l,m)} C_{N(l,m)+s}^{2 s}
\left( {|{\bf q}+{\bf G}| d\over 2}\right)^{m+l-2 s\over 2} \cr \times
\{ e^{- |{\bf q}+{\bf G}| d} [\Gamma(s+{1\over 2})-
{\rm sgn}(f_{-})
\gamma(s+{1\over 2},f_-^2)] +
(-1)^{l+m-2 s\over 2} e^{|{\bf q}+{\bf G}| d} \Gamma(s+{1\over 2},f_+^2)\},
\eqnum{A.16}
\end{eqnarray}
where
\begin{eqnarray}
f_{\pm}={|{\bf q}+{\bf G}|\over 2\sqrt{\pi n}}\pm\sqrt{\pi n}d,
\eqnum{A.17}
\end{eqnarray}
\begin{eqnarray}
N(l,m)={\rm max}({m-l\over 2},{l-m-2\over 2}).
\eqnum{A.18}
\end{eqnarray}

In deriving (A.16) we took into account that in the problem considered,
the parameters $l$ and $m$ have the same parity. The term with ${\bf
G} = 0$ in (A.16) is compensated by the interaction with the positive
background of the adjacent layer. Using (A.13), (A.14), and (A.16),
we can write the contribution of the interlayer interaction in the form
(16).

\end{document}